\documentclass[9 pt,a4paper]{article}
\usepackage[latin1]{inputenc}
\usepackage{amsmath}
\usepackage{amsfonts}
\usepackage{amssymb}
\title{An Energy Puzzle in Quantum Collapse}

\author{Thiago Guerreiro\footnote{thguerreiro@gmail.com}  and Fernando Monteiro\footnote{fhrmonteiro@gmail.com }\\
		\begin{small}
	Departamento  de F\'{\i}sica, Pontif\'{\i}cia Universidade Cat\'olica
\end{small}\\
		\begin{small}
Rio de Janeiro, Brazil
\end{small}}
\date{}
\begin{document}
\maketitle

\begin{abstract}
Since its discovery, quantum theory has proven to be one of the most precise theories ever made. Measurement processes, however, do not seem to be governed by the unitary law of quantum mechanics, and one can ask if the theory is complete. To answer this question, new experiments must be done regarding the nature of quantum measurements. Here, we propose a direct consequence of projective measurements, leading to three possibilities. An arbitrarily high amount of energy can be either extracted or absorbed, or else energy conservation cannot be accounted by usual arguments based upon standard quantum mechanics. 
\\
\end{abstract}

Quantum mechanics' greatest triumph is also the cause of its biggest problem. There are very strong experimental indications$^{1-3}$ that there should be no fundamental limits to the applicability of quantum theory. This success implies that everything should be described by wave functions following suitable unitary evolutions. However, it is well known that nature seems to follow a different path during the process of measurements$^{4-6}$. 
  
The so called collapse postulate of the wave function has been a subject of intense debate over the last decades, and it has been recognized that geometric conservation laws such as energy and momentum can be violated under the presence of this state reduction$^{7}$. With all the current advance in quantum information technology, and with quantum simulators everyday closer to becoming a reality$^{8, 9}$, can we study these violations in the laboratory, and use them to learn something new about measurements in the quantum realm? 

To probe this question we propose a toy model that mimics a measurement process exactly as the one described by Von Neumann's projection postulate$^{10}$. The validity of such paradigm is questionable and our scheme outlines a direct experimental consequence of it, related to violation of conservation laws. The observation of such experimental consequence would constitute an energy puzzle directly connected with the problem of measurements in quantum mechanics.

Let us assume that a system and detector, each represented by a qubit, interact and evolve according to
\begin{eqnarray}
\vert \Psi_{0} \rangle = \dfrac{1}{\sqrt{2}}\left( \varphi_{1} + \varphi_{2} \right) \otimes \psi_{0}  \rightarrow  \dfrac{1}{\sqrt{2}}\left(  \varphi_{1} \otimes \psi_{0} + i \varphi_{2} \otimes \psi_{1}  \right) = \vert \Psi \rangle
\end{eqnarray}
where the left hand side of the arrow corresponds to the total state at $ t = 0 $, and the right hand side at $ t = t_{1} $. 

Postulating that collapse as in the Copenhagen interpretation of quantum mechanics takes place at $ t_{1} $ with $ \psi_{0} $ and $ \psi_{1} $ as the pointer basis, a condition that must be satisfied for the expectation value of energy to be conserved is$^{8}$
\begin{eqnarray}
\langle \Psi \vert \hat{H} \vert \Psi \rangle  = \dfrac{1}{2}\left[ \langle \varphi_{1} \otimes \psi_{0} \vert \hat{H} \vert \varphi_{1} \otimes \psi_{0} \rangle + \langle \varphi_{2} \otimes \psi_{1}  \vert \hat{H} \vert \varphi_{2} \otimes \psi_{1} \rangle \right]
\end{eqnarray}
where $ \hat{H} $ is the Hamiltonian that generates the dynamics given by (1). We wish to construct a scheme that evolves according to $ \hat{H} $ and violates this conservation condition.  

Representing the total Hilbert space basis 
\begin{center}
$ \lbrace \varphi_{1} \otimes \psi_{0} , \varphi_{1} \otimes \psi_{1} , \varphi_{2} \otimes \psi_{0} , \varphi_{2} \otimes \psi_{1} \rbrace $
\end{center}
as a canonical ordered basis of $ C^{4} $, a possible choice of unitary that performs our desired evolution is 
\begin{eqnarray}
\hat{U}(t_{1}) = \left( \begin{array}{ccccc}
0 & 0 & 1 & 0 \\
0 & 0 & 0 & 1 \\
0 & -i & 0 & 0 \\
i & 0 & 0 & 0  \end{array} \right)
\end{eqnarray}

Since we have $ \exp \left( -\frac{i}{\hbar} \hat{H} t \right) = \hat{U}(t) $ we can write as a possible choice for Hamiltonian
\begin{eqnarray}
\hat{H} = \dfrac{\hbar}{i} \ln \hat{U}(t_{1}) = \dfrac{h}{8} \left( \begin{array}{ccccc}
1 & -i & -1+i & -1+i \\
i & 1 & 1-i & -1+i \\
-1-i & 1+i & 1 & -i \\
-1-i & -1-i & i & 1  \end{array} \right)
\end{eqnarray}
and $ \hat{U}(t) $ results in (1) for $ t_{1} = 1 + 4 k $, $ k = 0, 1, 2, ... $ It is easy to show that this Hamiltonian violates the conservation condition. The expectation value $  \langle \Psi \vert \hat{H} \vert \Psi \rangle  = 0 $, while
\begin{eqnarray*}
\dfrac{1}{2}\left[ \langle \varphi_{1} \otimes \psi_{0} \vert \hat{H} \vert \varphi_{1} \otimes \psi_{0} \rangle + \langle \varphi_{2} \otimes \psi_{1}  \vert \hat{H} \vert \varphi_{2} \otimes \psi_{1} \rangle \right] = \dfrac{h}{8}
\end{eqnarray*}

We can expand 
\begin{eqnarray*} 
 \langle \Psi \vert \hat{H} \vert \Psi \rangle  = \dfrac{1}{2}\left[ \langle \varphi_{1} \otimes \psi_{0} \vert \hat{H} \vert \varphi_{1} \otimes \psi_{0} \rangle + \langle \varphi_{2} \otimes \psi_{1}  \vert \hat{H} \vert \varphi_{2} \otimes \psi_{1} \rangle \right] + \\
 + \dfrac{i}{2} \left[ \langle \varphi_{1} \otimes \psi_{0}  \vert \hat{H} \vert \varphi_{2} \otimes \psi_{1} \rangle - \langle \varphi_{2} \otimes \psi_{1} \vert \hat{H} \vert \varphi_{1} \otimes \psi_{0} \rangle \right]
\end{eqnarray*}
and see that the cross terms in the above equation are responsible for this energy expectation difference. Thus, if Copenhagen is correct and measurements follow Von Neumann's projection postulate we gain energy on average after measuring the system's state with this scheme.

To better understand this example, let us try to decompose this Hamiltonian into something of the form $ \hat{H} = \hat{H}_{1} + \hat{H}_{2} + \hat{H}_{I} $. Representing
\begin{center} 
$ \varphi_{1} = \psi_{1} =  \left( \begin{array}{ccc}
1 \\
0  \end{array} \right) $ , $ \varphi_{2} = \psi_{2} =  \left( \begin{array}{ccc}
0 \\
1  \end{array} \right) $
\end{center}
we can see that our scheme is measuring $ \sigma_{z} $, and
\begin{eqnarray}
\hat{H} = \dfrac{h}{8}\left\lbrace  \left( -\sigma_{y} - \sigma_{x} + \hat{1} \right) \otimes \hat{1} + \hat{1} \otimes \sigma_{y} - \sigma_{x} \otimes \sigma_{y} + \sigma_{y} \otimes \sigma_{y} \right\rbrace 
\end{eqnarray}
where the two first terms are identified with $ H_{1} $ and $ H_{2} $ respectively, and the last two terms with $ H_{I} $.

An interesting fact about this example is that
\begin{eqnarray}
\exp \left[ -\frac{i}{\hbar} \hat{H} (3+4k) \right] = \left( \begin{array}{ccccc}
0 & 0 & 0 & -i \\
0 & 0 & i & 0 \\
1 & 0 & 0 & 0 \\
0 & 1 & 0 & 0  \end{array} \right)
\end{eqnarray}
so that at the instants $ \tilde{t}_{1} = 3 + 4k $, $ k = 0, 1, 2,... $ our system and detector evolve like
\begin{eqnarray}
\vert \Psi_{0} \rangle = \dfrac{1}{\sqrt{2}}\left( \varphi_{1} + \varphi_{2} \right) \otimes \psi_{0}  \rightarrow  \dfrac{1}{\sqrt{2}}\left( i \varphi_{1} \otimes \psi_{1} + \varphi_{2} \otimes \psi_{0}  \right) = \vert \tilde{\Psi} \rangle
\end{eqnarray}
This is very similar to (1), but with exchanged pointer positions. Even if collapse were to occur at any of the $ \tilde{t}_{1} = 3 + 4k $, $  \langle \tilde{\Psi} \vert \hat{H} \vert \tilde{\Psi} \rangle  = 0 $ and
\begin{eqnarray*}
\dfrac{1}{2}\left[ \langle \varphi_{1} \otimes \psi_{1} \vert \hat{H} \vert \varphi_{1} \otimes \psi_{1} \rangle + \langle \varphi_{2} \otimes \psi_{0}  \vert \hat{H} \vert \varphi_{2} \otimes \psi_{0} \rangle \right] = \dfrac{h}{8}
\end{eqnarray*}
so if we postulate state reduction at one of these instants, energy expectation conservation is still not respected. 

In the case that collapse happens at $ t_{1} $, the final state after measurement can be written as
\begin{eqnarray*}
\rho = \dfrac{1}{2} \vert \varphi_{1} \otimes \psi_{0} \rangle \langle \varphi_{1} \otimes \psi_{0} \vert + \dfrac{1}{2} \vert \varphi_{2} \otimes \psi_{1} \rangle  \langle \varphi_{2} \otimes \psi_{1}  \vert
\end{eqnarray*}

Let us now bring the state $ \rho $ back to $ \vert \Psi_{0} \rangle $ by any process involving the least possible energy difference. Then one of three possible conclusions has to be true. Either one can extract from the system plus detector an arbitrary amount of energy, or one can sink into it an arbitrary amount of energy, or there is an energy conservation law not deducible from the usual physical postulates. Each conclusion is interesting new physics.

To postulate that collapse should happen at $ t_{1}= 1 + 4k $ or at $ \tilde{t}_{1} = 3 + 4k $ is questionable. Why should state reduction take place at any of these times? These instants are the only ones for which orthogonal states of the detector are correlated with orthogonal states of the system in the basis we are working. So this leads to another question. 

Why should collapse occur with $ \psi_{0} $ and $ \psi_{1} $ as pointer basis? This is the famous problem of preferred basis. Since there is no widely accepted solution to both the preferred basis problem and the problem of definite outcomes$^{6}$, we argue that our example qualifies as a measurement scheme, in the sense put forward by Von Neumann, just as any other entangling scheme between system and detector. 

Another important point is that this example treats the detector as a qubit, having thus the same status as the system. A proper way of viewing and justifying this would be to assume that the states spanned by $ \psi_{0} $ and $ \psi_{1} $ correspond to a two dimensional degree of freedom of the \textit{true detector}, a macroscopic object whose actual state lives in $ \mathcal{H}^{N} $, where $ N \gg 10^{23} $. 

Since quantum mechanics appears to be as good for macroscopic objects as it is for subatomic particles, the correct way to write the initial total state would be $ \Psi_{0}(\varphi, \psi) \otimes \Lambda $ where $ \Lambda $ corresponds to the other degrees of freedom of the detector and the rest of the universe. The unitary evolution that correlates system and detector would lead to entanglement between the system's states $ \varphi_{1} $ and $ \varphi_{2} $, and a state in $ \mathcal{H}^{N} $.

Collapse would then have to be postulated, to account for our observations of definite outcomes, and by choosing an appropriate Hamiltonian $ H + H_{\Lambda} + H_{I, \Lambda} $ we could be led to the same conclusions as in our example. There is no fundamental difference between considering a two level system as detector and taking into account a whole $ \mathcal{H}^{N} $ macroscopic object if we assume that unitary quantum theory is universal. 

In a more realistic situation, the detector qubit could be treated as a ``quantum probe'' and the interaction with the rest be of the non demolition kind, such that $ \left[ H, H_{I, \Lambda} \right] = 0 $. This would certify that the system and quantum probe energy is a constant of the evolution until collapse takes place. After that, the system and quantum probe's expectation of energy would change due to state reduction, and would remain constant throughout the rest of the unitary evolution. 

A Hamiltonian with the same properties as $ \hat{H} $ could in principle be implemented in the laboratory using quantum computing technologies. A natural candidate for system and quantum probe would be superconducting qubits, due to their versatility$^{11, 12}$. Also, quantum non demolition measurements have been demonstrated using circuit quantum electrodynamics set ups$^{13}$.

We are led to two possibilities. Collapse is nothing but a postulate and that is as far as we can go. There can be no dynamics governing it, and it must be an axiom of the theory to account for our observation of definite outcomes. Under that possibility, Von Neumann's projection is correct and there are no a priori physical principles that forbids energy expectation values to be arbitrarily increased or decreased. The other possibility, collapse can be explained within the framework of new physics, and the energy expectation change has its origin on a new unknown theory, that extends quantum mechanics and reduces the collapse postulate to an effective description of measurements. Only future experiments can lead to such theory.
\\

\title{\textbf{Acknowledgement}}
\\

We wish to thank Professor G. Svetlichny, without whom this work could not be done, for stimulating discussions, comments and suggestions.
\\

\title{\textbf{References}}
\begin{enumerate}
\item L. Hackermuller \textit{et. al.} Wave nature of Biomolecules and Fluorofullerenes. \textit{Phys. Rev. Lett.} \textbf{91}, 090408 (2003).
\item Gerlich, S. et al. Quantum interference of large organic molecules. \textit{Nat. Commun.} \textbf{2}, 263 doi: 10.1038/ncomms1263 (2011).
\item O'Connell, A. \textit{et. al.} Quantum ground state and single-phonon control of a mechanical resonator. \textit{Nature} \textbf{464}, 697-703, 1 April 2010

\item Bassi, A. \textit{et.} Ghirardi, G. A general argument against the validity of the superposition principle. \textit{Phys. Lett. A} \textbf{275}, 373-381 (2000)

\item Adler, S. Why decoherence has not solved the measurement problem: A response to P. W. Anderson. \textit{Studies In History and Philosophy of Science Part B} \textbf{34}, 135-142 (2003)

\item Schlosshauer, M. Decoherence, the measurement problem and interpretations of quantum mechanics. \textit{Rev. Mod. Phys.} \textbf{76}, 1267-1305 (2005) 

\item Pearle, P. Wave-function collapse and conservation laws. \textit{Found. of Physics} \textbf{30}, 8, 1145-1160 (2000)

\item Buluta, I. \textit{et. al.} Quantum simulators. \textit{Science} \textbf{326}, 108 (2009)

\item Ladd, T. D. \textit{et. al.} Quantum computers. \textit{Nature} \textbf{464}, 45-53, 4 March 2010
\item J. von Neumann, Mathematisce Grundlagen der Quantummechanik, Springer, Berlin, 1932
\item Makhlin, Y. \textit{et. al.} Josephson-junction qubits with controlled couplings. \textit{Nature} \textbf{398}, 305-307 (1999)
\item Leek, P. J. \textit{et. al.} Observation of Berry's phase in a solid state qubit. \textit{Science} \textbf{318}, 1889-1892
\item Lupascu, A. \textit{et al.} Quantum non-demolition measurement of a superconducting two-level system. \textit{Nature Physics} 3, 119-125 (2007)

\end{enumerate}

\end{document}